# Dispersion-less Kerr solitons in spectrally confined optical cavities

Xiaoxiao Xue[1*], Philippe Grelu[2], Bofan Yang[1], Mian Wang[1], Shangyuan Li[1], Xiaoping Zheng[1], and Bingkun Zhou[1]

[1]*Department of Electronic Engineering, Beijing National Research Center for Information Science and Technology, Tsinghua University, Beijing 100084, China.*

[2]*Laboratoire ICB, UMR 6303 CNRS, Université Bourgogne Franche-Comté, 21000 Dijon, France.*

*[*xuexx@tsinghua.edu.cn](mailto:xuexx@tsinghua.edu.cn)*

## Abstract

Solitons are self-reinforcing localized wave packets arising from a balance of linear and nonlinear effects. This definition encompasses the interplay of nonlinear gain and loss, leading to the concept of dissipative solitons that has been instrumental in understanding the wide variety of mode locking phenomena in ultrafast optics. To date, most studies have involved the group velocity dispersion as a key ingredient for soliton generation. Here, we report on a novel kind of soliton, both theoretically and experimentally, which builds up in spectrally confined cavities when dispersion is practically absent. Precisely, the interplay between the Kerr nonlinearity and spectral filtering results in an infinite hierarchy of eigenfunctions which, combined with optical gain, allow for the generation of stable dispersion-less dissipative solitons in a previously uncharted regime. When the filter order tends to be infinite, we find an unexpected link between dissipative and conservative solitons, in the form of Nyquist-pulse-like solitons endowed with an ultra-flat spectrum. In contrast to the dispersion-enabled solitons, these dispersion-less Nyquist solitons build on a fully confined spectrum and their energy scaling is not constrained by the pulse duration. This study broadens the fundamental scope of dissipative soliton physics and opens new avenues for engineering optical solitons endowed with superior temporal and spectral features.

Solitons and solitary waves are ubiquitous in nature and widely studied across the disciplines of nonlinear science, such as mathematics, optics, condensed matter, chemistry, and biology [1]-[3]. The stationary pulsed solutions of the conservative nonlinear Schrödinger (NLS) equation that governs, for instance, the optical field propagation in fiber waveguides have attracted a broad interest from the early days of optical communications and fiber laser technology [4]. Indeed, the cancellation of the anomalous group velocity dispersion by the Kerr nonlinearity in shape-preserved transform-limited solitary pulses once appeared as a fascinating prospect for high-capacity optical communications. Soon afterwards, due to a growing interest in developing novel mode-locked laser architectures [5] and Kerr frequency combs [6] for metrology applications, the research was largely extended to the investigation of dissipative solitons in optical cavities. Dissipative solitons are sustained by the composite balance between dispersion and nonlinearity on the one hand and gain and loss on the other hand. The contribution of dissipation increases the soliton diversity well beyond that available in Hamiltonian conservative systems. For instance, the mode locking of bright pulses in the normal dispersion regime becomes possible with the help of spectral filtering. This led to remarkable experiments highlighting the superior pulse energy scaling in such regime [7]-[9]. In contrast to the conservative NLS solitons, dissipative solitons feature frequency chirping, which reflects a necessary energy redistribution within the pulse as it propagates. Whatever the dispersion regime, the group velocity dispersion remains an essential physical effect affecting the pulse existence in the space of cavity parameters and influencing strongly the optical pulse shaping. Although the existence of solitons in the absence of chromatic dispersion has been suggested [10], no clear supportive results have ever been reported so far. Exploring such possibility is not only important for a broader understanding of the fundamental nonlinear dynamics of complex systems, but it is likely to enable novel ultrafast sources for a wide range of applications.

In this article, we present our findings concerning a novel category of dissipative optical solitons, which we call *dispersion-less solitons* as they arise from the interplay between dissipation and Kerr nonlinearity in the absence of group velocity dispersion. We base these findings on a well-known physical system, suitable for both theoretical modeling and experimental testing: the optically driven Kerr cavity [11]-[13]. Driving the Kerr cavity to new parameter regions, we uncover previously uncharted regimes in which ultra-flat spectrum, transform-limited pulse generation and superior energy scaling could be simultaneously achieved. To do so, we precisely tailor the spectral filtering applied to the intracavity field while cancelling the path-averaged chromatic dispersion. This leads to new stationary regimes that are analyzed numerically and tested experimentally in a fiber ring cavity. In the following, we show that these dispersion-less solitons are closely related to the eigenfunctions of combined self-phase modulation (SPM) and spectral filtering effects, which asymptotically evolve toward an energy-conserved pulsed solution with the increase of filter order. Such remarkable trend thus reveals the existence of Kerr-only conservative solitary waves as limiting cases of the dissipative dispersion-less solitons when the spectral filtering becomes an ideal bandpass transmission. By construct, the asymptotic solitary wave shares the properties of an ideal Nyquist pulse, featuring a fully confined ultra-flat spectrum and a transform-limited temporal waveform [14]-[17]. Owing to their flat and compact spectral support, Nyquist pulses would be instrumental in improving the flatness of optical frequency comb, thus benefitting many applications such as high-capacity telecommunications [17], [18], spectroscopy [19], optical neural computing [20], [21], and radiofrequency photonics [22]. However, their direct synthesis as self-generated soliton pulses remain particularly challenging [14], [15]. In our experimental validation, by pumping a Kerr fiber cavity incorporating programmable dispersion and loss, we highlight the transformation of the dispersion-less soliton profiles when the filter order is increased. At high filter order, we demonstrate the deterministic generation of a Nyquist-pulse-like soliton endowed with a flat spectrum and a transform-limited temporal waveform.

**Principles to generate dispersion-less cavity solitons**

Our system model is that of the optically driven Kerr ring cavity, which has stimulated numerous theoretical investigations and yielded remarkable experimental advances with an emphasis on its suitability to produce coherent optical frequency combs through the generation of dissipative Kerr cavity solitons [6]. According to its implementation footprint, we can draw two main categories of experimental platforms that are conceptually equivalent but display important practical differences. The microresonator platform is targeting on-board chip-scale integration and naturally provides a multi-GHz spacing between comb lines that is suitable for major applicative needs. The optical fiber cavity platform involves a macroscopic setup allowing an easier management of the physical effects occurring in the field propagation, such as dispersion management or loss compensation schemes, through the insertion of specific optical components [9], [23]-[25]. To demonstrate novel cavity soliton solutions, we need to precisely control the chromatic dispersion, the gain/loss balance and the intracavity spectral filtering. Therefore, our chosen experimental platform is the optical fiber cavity, nevertheless the conceptual advances remain applicable to the microresonator platform through a proper scaling and design of the photonic waveguide structures. Let us give an example of transposition between both platforms that concerns the control of spectral filtering, an important feature of the present work. Whereas in fiber cavities, it can be easily implemented by inserting a discrete optical filter element, photonic crystal structures can be designed to provide a specific spectral transmission and fabricated within integrated platforms [26]. As such, Figure 1a shows a conceptual design of an integrated ring cavity that can be coherently driven to sustain dissipative solitons for Kerr frequency comb generation. Spectral control is integrated through a periodic Bragg boundary that is implemented on the inner side of the ring waveguide. Therefore, the frequency components falling within the stopband can be confined in the cavity while the out-of-band frequencies experience high loss and will leak out. To search the dispersion-less soliton solutions that form when the average group velocity dispersion becomes negligible, we use the following normalized distributed propagation model (see the Supplementary Section 1.1 for its derivation):

$$\frac{\partial U}{\partial Z} = i|U|^2 U - \left(\frac{i}{\pi}\frac{\partial}{\partial T}\right)^n U - i\Delta U - U + \left[iC^* U^2 + iC^2 U^* + i2C|U|^2\right] + i2|C|^2 U \qquad (1)$$

where $U$ is the field envelope of the optical pulse; $Z$ and $T$ are the scaled spatial and temporal coordinates; $\Delta$ is the pump-cavity phase detuning; $C$ is the homogenous background of the pumped mode in the cavity; $n$ is the filter order. Here, the normalized filter bandwidth out of which the spectral loss exceeds the uniform loss is 1. With the filter order $n \to \infty$, the spectrum of $U$ becomes fully confined in a unit frequency span. The three terms enclosed in square brackets represent the parametric mixing between the soliton and the background pump. The last term is an additional phase shift induced by cross-phase modulation that is usually negligible in comparison to $\Delta$.

To gain more insight into the dispersion-less soliton dynamics and how the balance can be achieved between dissipation and Kerr nonlinearity, the following eigenvalue equation is investigated:

$$i|U_e|^2 U_e - \left(\frac{i}{\pi}\frac{\partial}{\partial T}\right)^n U_e = (i\xi + \lambda)U_e \qquad (2)$$

where $U_e$ is the eigenfunction of combined SPM and spectral filtering; $\xi$ and $\lambda$ are the imaginary and real parts of the eigenvalue respectively. It turns out that $\lambda$ is always negative: this reflects an overall amplitude loss induced by spectral filtering, whose impact also depends on SPM since the latter is conducive to the broadening of the optical spectrum. Apparently, these eigenfunction solutions are not stationary solutions of the propagation equation (1) but represent pulses damped with a decay rate contribution from spectral filtering and SPM that is given by $|\lambda|$. However, when the pulses are coherently pumped, the energy loss can be compensated by the parametric gain, so that a composite balance can be achieved between all these propagation effects, resulting in the formation of cavity solitons. In the schematic representation of such composite balance by counteracting arrows in Fig. 1b, the vertical dimension represents purely dissipative effects and the horizontal one, purely dispersive contributions. Therefore, the tilt of the parametric gain arrow reflects its frequency dependence, which contributes to the balancing of SPM in absence of chromatic dispersion. More remarkably, we find that for a given value of $\xi$, the energy loss induced by spectral filtering decreases and tends asymptotically to zero with the increase of filter order $n \to \infty$ (see Supplementary Section 1.2). Therefore, in the limit when the filter order tends to infinity, spectral filtering and SPM can balance each other, leading to an eigenfunction that becomes a Kerr-only solitary wave in an ideal bandpass system. In a realistic system, though, we should include linear losses, which can be precisely compensated by the parametric gain.

Figure 1c displays the numerically solved results of Eqs. (1) and (2) with $\xi = \Delta = 20$, at different filter orders $n$. When the filter order increases, the soliton spectrum becomes more confined, and the pulse chirp decreases. When the filter order $n \to \infty$, the soliton spectrum becomes fully confined within a unit bandwidth, featuring an intensity variation limited here to 4.2 dB. We also see that the asymptotic soliton adopts a uniform spectral phase, which is perfectly consistent with the fact that, as the losses vanish, the soliton behaves as a stationary conservative soliton. Correspondingly, the soliton temporal waveform becomes a transform-limited Nyquist pulse with characteristic oscillating tails. An approximate closed-form solution can be derived for the Nyquist soliton with an ansatz method. A perturbation analysis of the governing averaged propagation model is performed in the Supplementary Section 1.3, showing that such eigenfunction is a stable attractor for coherently driven Kerr cavities with ideal bandpass transmission and negligible chromatic dispersion.

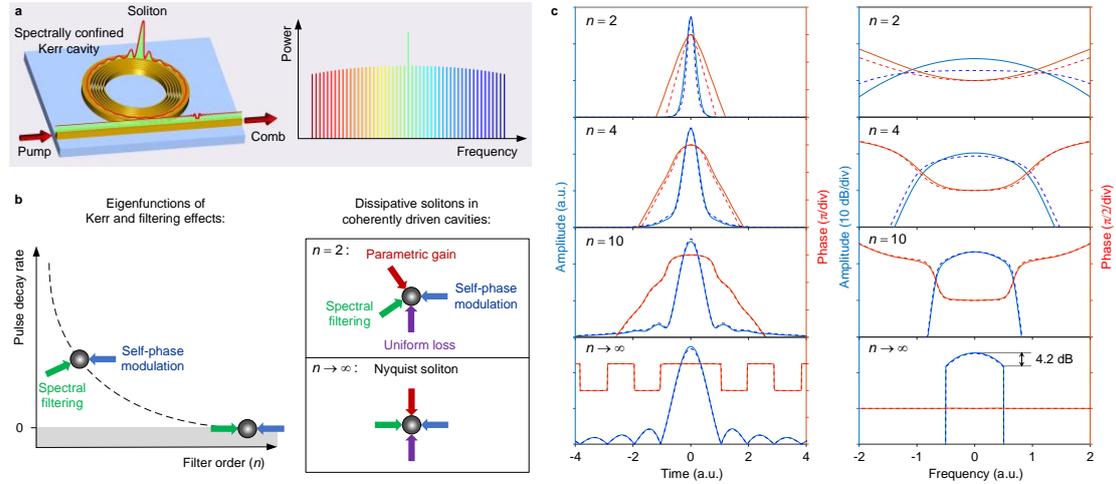

**Fig. 1 | Principle of the dispersion-less cavity soliton. a,** A conceptual design of spectrally confined ring cavity for Kerr frequency comb generation. The inner side of the ring waveguide is formed by a band-stop Bragg boundary. **b,** Schematic of the Kerr-and-filter eigenfunctions (left) and the dispersion-less stationary cavity solitons (right). The Kerr-and-filter eigenfunctions have a positive decay rate, which tends to zero when the filter order tends to infinity. Within a coherently pumped cavity, the parametric gain and uniform loss can achieve a composite balance to sustain dispersion-less cavity solitons whatever the filter order. When the filter order $n \to \infty$, a dispersion-less Nyquist soliton results from the double balance between spectral filtering and self-phase modulation on the first hand, and parametric gain and uniform loss on the second hand. **c,** Numerical results of the Kerr-and-filter eigenfunctions (dashed lines) and the cavity solitons (solid lines). Left, temporal amplitude and phase; right, spectral amplitude and phase.

## Results

**Observation of dispersion-less cavity solitons.** A fiber ring cavity comprising single-mode highly nonlinear as well as standard optical fibers is built to study the soliton dynamics. The experimental setup is shown in Fig. 2a. A spectral shaper is inserted for programmable dispersion and filter control [27], [28], and a short length of erbium-doped fiber (EDF) is used to compensate for the roundtrip loss. The EDF length, its optical pumping power, and the pumping strength of the ring cavity are carefully tailored such that the gain provided by the optical amplifier keeps the ring cavity below laser threshold, and that the optical amplifier operates in the linear regime. Therefore, the behavior of the setup can be characterized as a coherently driven passive cavity for which the effective roundtrip loss is given by the net loss of the fiber loop [25]. Figure 2b shows the through-port transmission of the cavity. The total roundtrip length is 54.4 m and the power loss is 16.6%, corresponding to a free spectral range (FSR) of 3.73 MHz and a finesse of 36. Various spectral filtering functions can be implemented by programming the amplitude of the spectral shaper, while the fiber group velocity dispersion is carefully compensated by programing the phase. Figure 2c shows an example of a 250-GHz gate bandpass filter. The spectral shaper has a limited optical resolution of 10 GHz, correspondingly, a maximum filter order of 30 can be achieved.

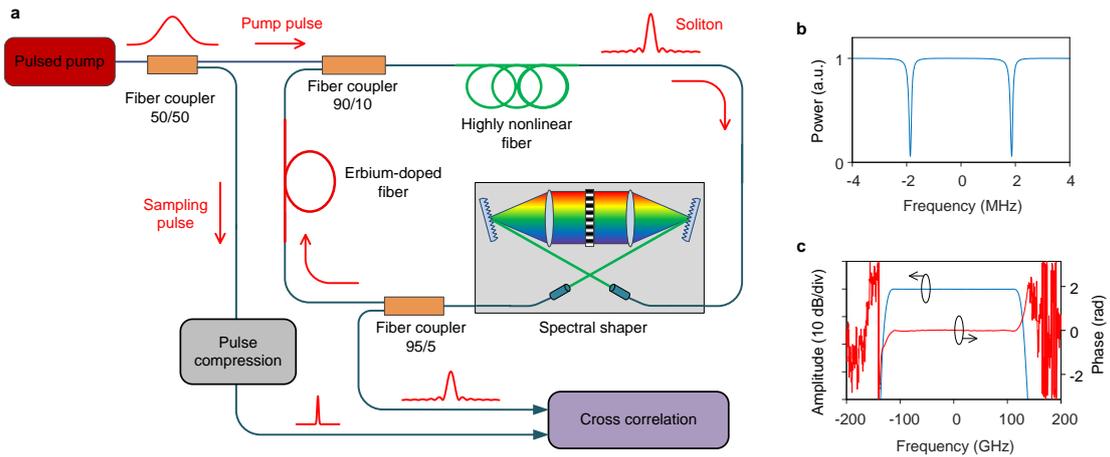

**Fig. 2 | Fiber ring cavity for studying the soliton dynamics. a,** Experimental setup. A spectral shaper is inserted in the cavity for programmable dispersion and filter control, and a section of erbium-doped fiber is used to compensate for the roundtrip losses. The cavity is synchronously pumped by a sequence of sub-nanosecond Gaussian pulses for soliton generation. A fraction of each pump pulse is spectrally broadened and compressed to sub-picosecond duration for measuring the soliton waveform via intensity cross correlation. **b,** Through-port response of the cavity measured with a low-power continuous-wave laser. **c,** Roundtrip transfer function of the cavity when the spectral shaper is programmed to a gate bandpass filter and the fiber group velocity dispersion is compensated.

The fiber ring is synchronously pumped by a sequence of 87.4-ps Gaussian pulses with a peak power of 1.5 W through a 90/10 coupler. The generated soliton is extracted from the 5% output of a 95/5 coupler for ultrafast characterization. To measure the soliton waveform, a high-resolution intensity cross-correlation setup is built, with the sampling pulses (0.2 ps) generated by spectrally broadening and compressing the pump pulses (more details presented in Supplementary Section 2). Figure 3 shows the results of the dispersion-less cavity solitons observed when the pump is tuned into the cavity resonance from the blue side (i.e., pump frequency higher than the resonant frequency). The spectral shaper is programmed to apply super-Gaussian filters with orders $n = 2$, 4, 10 and 30, respectively. The filter bandwidth is 250 GHz. An excellent agreement between the experiment and the simulation is obtained. With the increase of the filter order, the soliton spectrum becomes rectangular, and the soliton temporal waveform gets closer to a Nyquist pulse. With the filter order $n = 2$, the measured soliton pulse full width at half maximum (FWHM) is about 5.5 ps while the transform-limited pulse duration assuming a uniform spectral phase would be 3.3 ps, entailing a pulse chirp parameter close to 1. In comparison, the soliton for $n = 30$ is nearly chirp-free with a FWHM of 3.7 ps and becomes close to a Nyquist pulse. For such Nyquist-like soliton, about 99.6% of the total spectral power is confined within the 250-GHz frequency window. The latter contains more than 66,000 comb lines within a 6-dB intensity range excluding the pump, corresponding to an unprecedentedly high spectral quality for Kerr frequency combs.

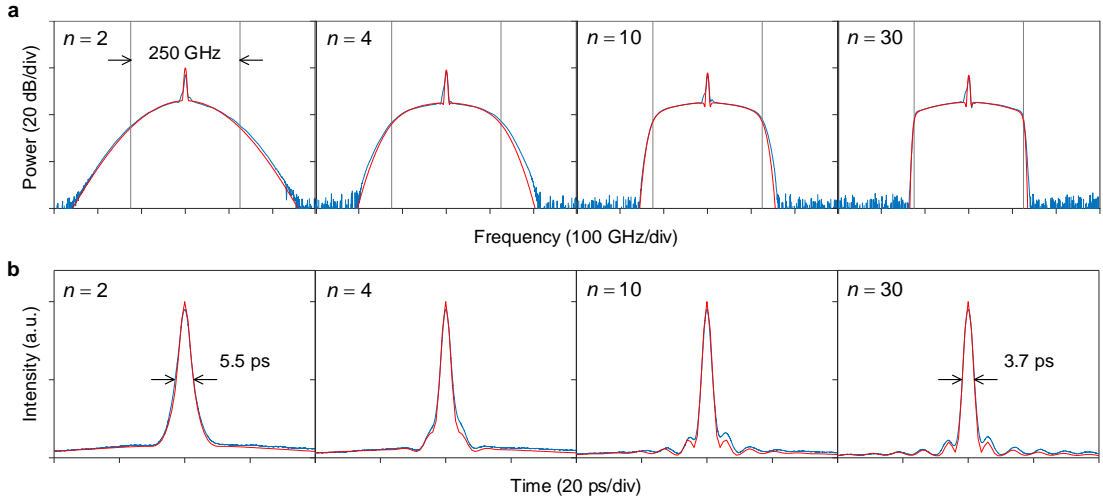

**Fig. 3 | Results of dispersion-less cavity solitons. a,** Optical spectrum. **b,** Temporal intensity waveform. Blue: measured; red: simulated. The filter bandwidth is 250 GHz, and the filter order $n$ = 2, 4, 10 and 30, respectively. With the increase of filter order, the soliton spectrum becomes more confined in the 250-GHz frequency window, and the soliton waveform tends to become a Nyquist pulse. The pulse width (FWHM) is 5.5 ps and the chirp parameter is 1 when $n$ = 2. The Nyquist soliton when $n$ = 30 is nearly chirp-free with a width of 3.7 ps.

**Nyquist soliton transition dynamics and its existence domain.** Using the maximum filter order $n = 30$, we find that the Nyquist soliton can be generated deterministically under pulsed pumping. Figure 4a shows the mean intracavity power trace when the pump laser scans across the resonance from the blue side. Steps corresponding to soliton state transitions can be clearly observed. In contrast to conventional dispersion-enabled cavity solitons which are usually excited stochastically through chaotic modulational instability [13], the Nyquist soliton generation is rather deterministic with an absence of noisy region observed from the intracavity power trace. The soliton waveform and spectrum at each step are measured. Some selected results are shown in Figure 4b (see Supplementary Section 2.3 for the full data). The solitons appear as square pulses with characteristic oscillating peaks and spectral sidelobes. The pulse duration is quantized and has a very good linear relation with the power step number (Figure 4c). After each transition step, the pulse width decreases by about 8.1 ps. Numerical simulations reveal that these square pulses are associated with compact Nyquist soliton molecules (i.e., closely bound solitons [5], [29]) which can also be sustained in ideal bandpass systems by the balance between Kerr nonlinearity and spectral filtering alone (Supplementary Section 1.4). With the increase of pump frequency detuning, the soliton number in the molecule decreases one by one and a single Nyquist soliton is ultimately formed. It is found that the soliton transition process can be affected by the desynchronization between the pump pulse repetition rate and the cavity FSR (Supplementary Section 2.4). In some cases, the existence region of some intermediate states may be too narrow to be observed and the soliton number decreases by two or more in a single transition step.

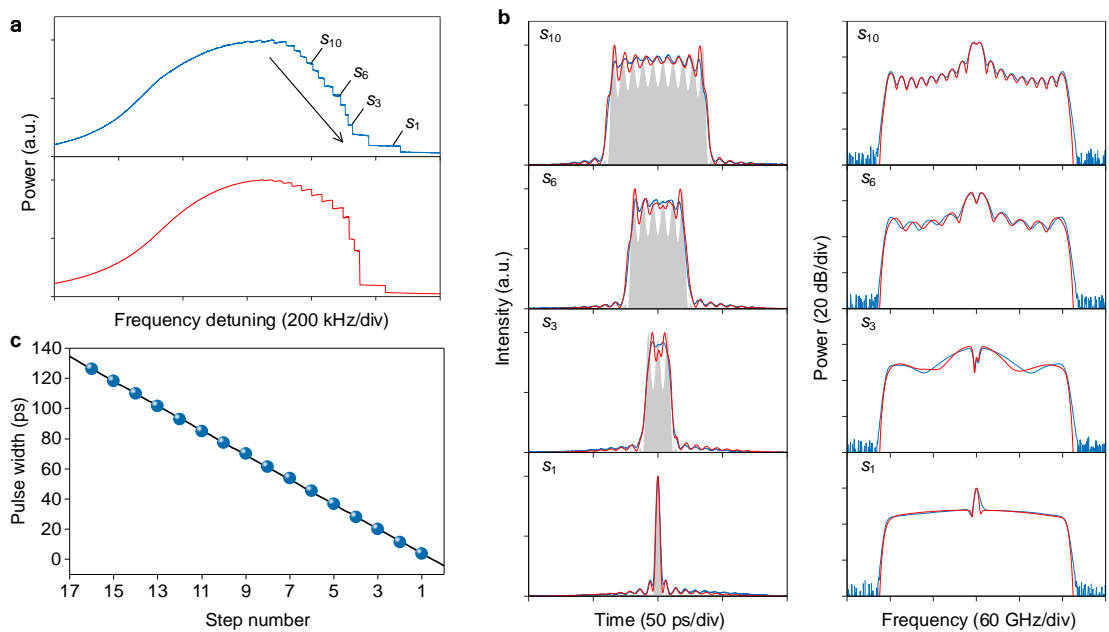

**Fig. 4 | Nyquist soliton transition dynamics under pulsed pumping. a,** Intracavity power versus pump frequency detuning (upper: measured; lower: simulated). **b,** Pulse optical intensity (left) and spectrum (right) at different steps marked in **a** (blue: measured; red: simulated). The gray shadows indicate the simulated soliton molecules sustained by a balance between Kerr nonlinearity and spectral filtering alone. With the increase of pump detuning, the soliton number decreases one by one, until a single Nyquist soliton is formed at the last step. **c,** Total width of the soliton molecule at each step. Blue dots: measured; black line: linear fit.

To explore the existence region of the single Nyquist soliton, a parameter scanning is performed numerically based on the normalized continuously-driven equation. The results are shown in Figure 5a. The Nyquist soliton can be maintained in a relatively wide parameter space that lies within the homogenously bistable region. Figure 5b shows the soliton pulse shape and spectrum at some selected locations marked in Fig. 5a. Along the lower boundary of the pump power (the minimum pump power required to maintain the soliton), the soliton intensity increases linearly with the increase of pump detuning. Meanwhile, the pulse width keeps nearly unchanged since the spectral bandwidth remains fully confined by the gate filter in the frequency domain. Detailed plots of the pulse energy and width are shown in Fig. 5c. The energy growing slope is about 1.5 and the pulse width is 0.94. This indicates that the ideal Nyquist soliton is free from energy-width scaling, in sharp contrast to the conventional solitons enabled by the second-order dispersion for which the energy scales with the inverse of the pulse duration. Therefore, the Nyquist soliton can in theory be promoted indefinitely while keeping almost the same pulse and spectral shape (more discussions about the scaling law of dispersion-less solitons are presented in Supplementary Section 1.2.2). When the pump detuning is kept constant, the soliton energy also increases slowly with the pump power, with an average slope of 0.035. The soliton oscillating tails get more prominent while the main pulse becomes slightly narrower. In the frequency domain, the spectral components close to the passband edge get more enhanced, giving rise to an increasingly flat spectrum. Along the dash-dotted line in Fig. 5a for which the pump phase detuning is kept at 20, the best achievable spectral flatness is about 2.3 dB when the pump power varies between the lower and upper boundaries.

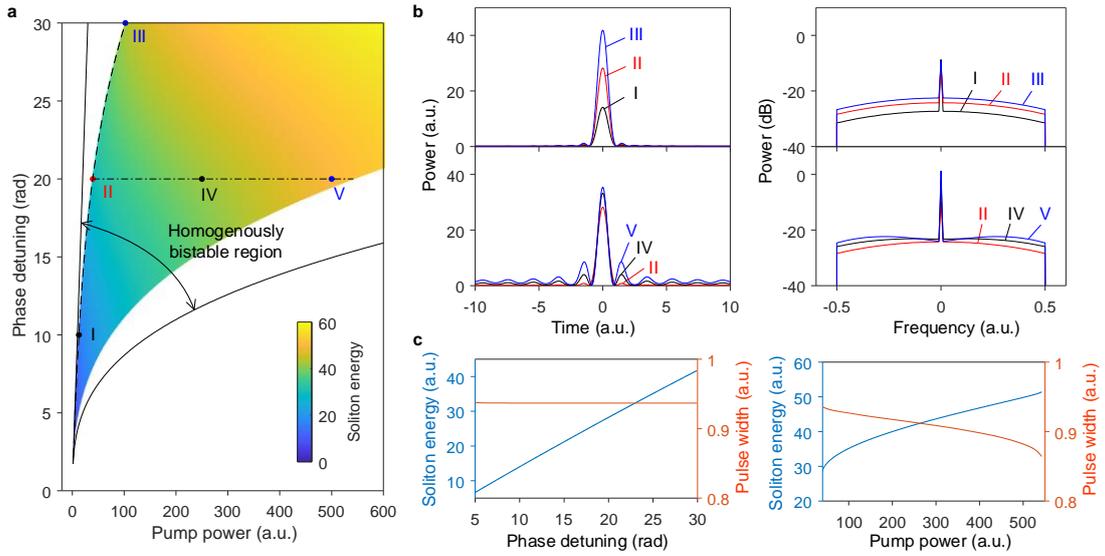

**Fig. 5 | Evolution of Nyquist solitons. a,** Soliton existence region. The dispersion-less Nyquist soliton can be maintained in a wide parameter space within the homogenously bistable region. **b,** Soliton pulse shape (left) and spectrum (right) at different locations marked in **a**. For I, II, and III, the pump phase detuning is 10, 20, and 30, respectively. For II, IV, and V, the pump phase detuning is 20 and the pump power is 40, 250, and 500, respectively. **c,** Plots of the pulse energy and width versus the pump phase detuning along the dash line in **a** (left), and versus the pump power along the dash-dotted line in **a** (right).

**Relation between filter-driven and dispersion-driven solitons.** When the net group velocity dispersion is not zero in the fiber loop, both the dispersion and the spectral filter will be responsible for soliton pulse shaping. Figure 6 shows the results when the spectral shaper is programmed to a 250-GHz gate filter and different amounts of second-order dispersion are applied. The soliton spectrum becomes slightly flatter when the net roundtrip dispersion is slightly normal (0.63 ps$^2$). However, the soliton existence region becomes narrower, and it is more difficult to get stable single Nyquist soliton when the dispersion is further increased. When the net dispersion is anomalous, the spectral intensity around the pump wavelength gets more pronounced. Two cases with net dispersion values of −1.06 ps$^2$ and −4.22 ps$^2$ are shown. With an increase of the net anomalous dispersion, the soliton tends to adopt hyperbolic-secant spectrum that is typical from conventional dispersion-driven solitons. The oscillating tails aside the main pulse, which are a signature of a Nyquist pulse, get diminished. A chaotic region with higher noise can also be observed from the intracavity power transition trace, which is attributed to the modulational instability in the anomalous dispersion region [13]. For all the three cases shown in Fig. 6b, the solitons are very close to transform-limited pulses with no obvious chirp.

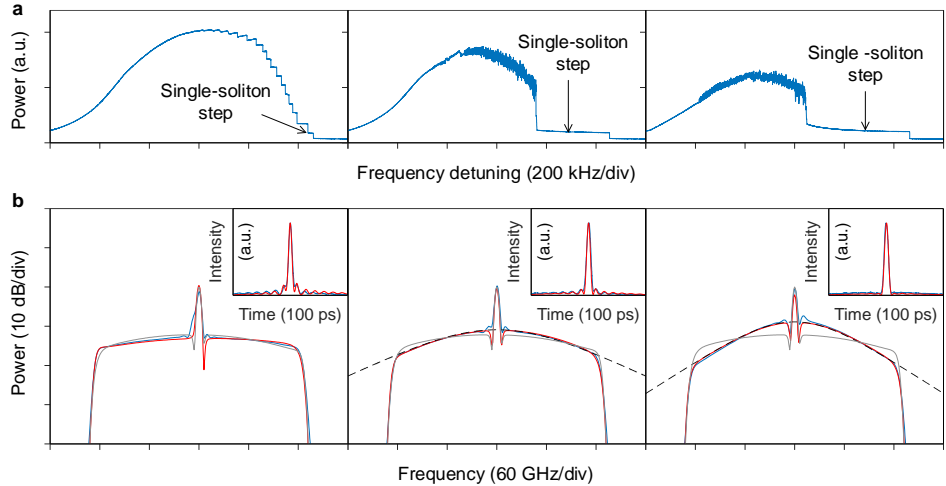

**Fig. 6 | Relation between filter-driven and dispersion-driven solitons. a,** Intracavity power versus pump frequency detuning. From left to right, the net roundtrip dispersion is 0.63 ps$^2$, -1.06 ps$^2$ and -4.22 ps$^2$, respectively. A chaotic region with higher noise can be observed when the dispersion is negative (anomalous). **b,** Soliton spectrum and pulse shape (inset) at the corresponding single-soliton steps indicated in **a** (blue: measured; red: simulated). The result with no dispersion is also shown for comparison (gray solid line). With the increase of anomalous dispersion, the pulse transitions from a filter-driven Nyquist soliton to a conventional dispersion-driven soliton for which the spectral envelope can be well fitted by a hyperbolic secant function (black dash line).

## Summary and discussion

In summary, this article demonstrates the existence of dispersion-less cavity solitons sustained by the balance between dissipation and Kerr nonlinearity. In addition to deepening the dissipative soliton paradigm, our study demonstrates an unexpected connection between dissipative and conservative cavity solitons by virtue of Nyquist pulses. We have developed a theoretical model relating the solitons to the eigenfunctions of self-phase modulation and spectral filtering effects. When the filter order tends to infinity, a Nyquist-pulse-like soliton is demonstrated. Such Nyquist soliton possesses several outstanding merits, including a fully confined flat spectrum, a transform-limited pulse shape, deterministic generation under pulsed pumping, and immunity from energy-width scaling. The soliton dynamics presented here can be exploited for frequency comb generation within miniature ring microcavities [6]. With appropriate design, it is also applicable to linear cavities, noting the recent upgrade in Kerr comb generation using Fabry–Pérot resonators formed by photonic crystal reflectors [26].

    We note that the dispersion-less solitons presented here are distinct from the "zero-dispersion" solitons that have been recently demonstrated in mode-locked lasers [28] and driven passive cavities [30] when the averaged second-order dispersion is vanishing. Indeed, the formation of solitons in [28], [30] is supported by the interaction involving the higher-order group velocity dispersion terms along with the Kerr nonlinearity. Moreover, whereas our experimental fiber ring is constructed from discrete elements with alternating normal and anomalous dispersion, the soliton dynamics is fundamentally different from the stretched-pulse mode-locking reported in [24], [31]. Such statement is supported by a comparison between the soliton spectra measured before and after the spectral shaper in our setup, see the Supplementary Section 2.5, showing no significant difference. This is explained by the fact that the local dispersive length is always much larger than the cavity length for the pulse duration involved, justifying the prevalence of the path-averaged cavity

dynamics. In our dispersion-less solitons experiments, we do not observe the Gordon-Kelly resonant sidebands that cause instabilities of high-energy pulses in dispersion-driven mode-locked fiber lasers [28], [32]. Since the group velocity dispersion is absent, the formation mechanism of Gordon-Kelly sidebands, namely constructive interference between the soliton and the dispersive wave, is no longer applicable. More interesting behaviors are worth further investigation for such pure-dissipation-enabled solitons in comparison to the conventional solitons.

Finally, whereas our experiments and theoretical analysis have been primarily conducted for coherently driven cavities, dispersion-less soliton regimes can also be engineered in mode-locked lasers. The numerical results of Nyquist pulse generation employing a mode-locked laser in [14] are likely related to the Nyquist soliton dynamics discovered here. We expect our findings to have a wide impact on soliton-based applications as well as in the understanding of complex nonlinear science.

**Methods**

**Soliton generation and characterization.** The fiber ring cavity is composed of a 21-m highly nonlinear fiber (YOFC NL-1550-NEG) and a ~30-m standard single-mode fiber. A commercial spectral shaper (Finisar) is inserted for programmable dispersion and filter control. A length of 0.6-m erbium-doped fiber (EDF, YOFC EDF13) is used to compensate for the loss. The EDF is pumped by a 1.8-W 976-nm laser, offering a gain of ~4.1 dB. The pulsed pump is generated by modulating a continuous-wave narrow-linewidth laser (OEwaves) with the signals from a pulse pattern generator (Anritsu), and is amplified to a peak power of 1.5 W for soliton sustainment. The amplified spontaneous emission noise is rejected by a narrow-band filter. To generate solitons, the pump laser frequency is manually tuned into the cavity resonance from the higher-frequency side. In the meanwhile, the relative detuning is stabilized by a home-built feedback servo controller to mitigate environmental interference. The soliton spectrum is measured by a commercial spectrum analyzer (Yokogawa) with a resolution of 0.02 nm; and the waveform is measured by a home-built intensity cross correlator having a resolution of 0.2 ps.

**Numerical simulations.** The results in Fig. 1c are obtained by numerically finding the solutions of Eqs. (1) (with the left side set to zero) and (2) with the Newton-Rapson method. The phase shifting parameters are $\xi = \Delta = 20$. Note that Eq. (1) is identical to the following equation

$$\frac{\partial \psi}{\partial Z} = -(1+\mathrm{i}\Delta)\psi + \mathrm{i}|\psi|^2 \psi - \left(\frac{\mathrm{i}}{\pi}\frac{\partial}{\partial T}\right)^n \psi + S \qquad (3)$$

where $\psi(Z,T) = U(Z,T) + C$ the total intracavity field; and $S$ is the pump field. In each case, the pump power $|S|^2$ is adjusted near its minimum value that is required to maintain the stable soliton. For the filter order $n = 2$, 4, 10 and $\infty$, the pump power $|S|^2 = 420$, 280, 120 and 25, respectively. The homogenous background $C$ is calculated by finding the lower-branch solution of Eq. (3) with the derivative terms set to zero.

When solving the eigenfunctions of Eq. (2), the real part of the eigenvalue (i.e., $\lambda$) can also be obtained at the same time as an unknown variable that depends on $\xi$.

The parameter scanning results in Fig. 5 are obtained by numerically integrating Eq. (3) with the standard split-step Fourier method until the intracavity field reaches a stable state. The cavity roundtrip time is 100. Note that the plots of Fig. 5b include both the soliton and the homogenous pump background, different from Fig. 1c which shows only the soliton.

The simulation results in Figs. 3, 4 and 6 for the fiber cavity are obtained with the split-step Fourier method, based on a simplified model in which the effects of Kerr nonlinearity, dispersion and loss are considered separately by lumped terms in one roundtrip. The field at the end of the ($m$+1)-th round is related to that after the $m$-th round by

$$A_{m+1} = \sqrt{(1-\theta)(1-\alpha_L)}e^{-i(\delta_L+\gamma P_m L)}FA_m + \sqrt{\theta}A_p \qquad (4)$$

where $A$ is the field amplitude normalized such that $P=|A|^2$ represents the field power; $L$ is the roundtrip length; $\alpha_L$ is the universal roundtrip power loss for all the modes; $\delta_L$ is the pump-cavity phase detuning; $\gamma$ is the average Kerr coefficient; $\theta$ is the pump-cavity power coupling ratio; and $A_p$ is the pump field. The operator $F$ accounts for the effects that can be easily applied in the frequency domain, including the group velocity dispersion, the frequency-dependent filtering loss, and the pump-cavity desynchronization. The simulation parameters are $\theta=10.4\%$, $\alpha_L=16.6\%$, $\gamma=4.6\times10^{-3}$ m$^{-1}$W$^{-1}$, and $L=54.6$ m. The pump pulse is Gaussian with a FWHM of 87.4 ps and a peak power of 1.5 W. The power transfer function of the super-Gaussian filter implemented by the spectral shaper is given by

$$|H(\nu)|^2 = e^{-[2(\nu-\nu_0)/B]^n} \qquad (5)$$

where $\nu_0$, $B$ and $n$ represent the filter central frequency, bandwidth and order, respectively.

## Acknowledgements

This work was supported by the National Key R&D Program of China (2018YFA0701902), the National Natural Science Foundation of China (61690192), and Zhejiang Lab (2020LC0AD01). Ph.G. acknowledges support from the EiPhi Graduate School (ANR-17-EURE-0004) and the French ISITE-BFC programs (ANR-15-IDEX-0003).

## Author Contributions

X.X. conceived the idea, and performed the experiments with help from B.Y., M.W. and S.L. X.X. and Ph.G. analyzed the results, performed the numerical simulations, and prepared the manuscript. All the authors are closely involved in discussions and revising the manuscript.

# Supplementary information to

# Dispersion-less Kerr solitons in spectrally confined optical cavities

Xiaoxiao Xue[1*], Philippe Grelu[2], Bofan Yang[1], Mian Wang[1], Shangyuan Li[1], Xiaoping Zheng[1], and Bingkun Zhou[1]

[1]*Department of Electronic Engineering, Beijing National Research Center for Information Science and Technology, Tsinghua University, Beijing 100084, China.*
[2]*Laboratoire ICB UMR 6303 CNRS, Université Bourgogne Franche-Comté, 21000 Dijon, France.*
*xuexx@tsinghua.edu.cn

## 1. Theoretical model

### 1.1 Mean-field equation for dispersion-less cavity solitons

The nonlinear dynamics in driven Kerr cavities with spectral losses was investigated before in [S1]-[S3]. When only the *n*-th order spectral loss is considered, the field evolution can be described by the following mean-field equation

$$\frac{\partial A}{\partial z} = -(\alpha + \mathrm{i}\delta)A + \mathrm{i}\gamma|A|^2 A - \rho\left(\mathrm{i}\frac{\partial}{\partial t}\right)^n A + \kappa A_\mathrm{p}, \tag{S1}$$

where $A$ is the field envelop; $t$ is the time; $z$ is the propagation distance; $\alpha$ is the uniform loss; $\delta$ is the pump-cavity phase detuning; $\gamma$ is the Kerr nonlinear coefficient; $\rho$ is the spectral loss; $A_\mathrm{p}$ is the pump field; and $\kappa$ is the pump coupling ratio.

The normalized form is given by

$$\frac{\partial \psi}{\partial Z} = -(1+\mathrm{i}\Delta)\psi + \mathrm{i}|\psi|^2 \psi - \left(\frac{\mathrm{i}}{\pi}\frac{\partial}{\partial T}\right)^n \psi + S, \tag{S2}$$

with

$$Z = \alpha z,\ T = Bt,\ \psi = AB\sqrt{\frac{\gamma}{\alpha}},\ \Delta = \frac{\delta}{\alpha},\ S = \frac{\kappa B A_\mathrm{p}}{\alpha}\sqrt{\frac{\gamma}{\alpha}},\ B = \frac{1}{\pi}\left(\frac{\alpha}{\rho}\right)^{1/n}. \tag{S3}$$

Suppose the solution is a soliton ($U$) sitting atop of a homogenous background ($C$)

$$\psi(Z,T) = U(Z,T) + C. \tag{S4}$$

Substituting Eq. (S4) into Eq. (S2), we get the equations for $U$ and $C$ respectively as follows

$$-(1+\mathrm{i}\Delta)C + \mathrm{i}|C|^2 C + S = 0, \tag{S5}$$

$$\frac{\partial U}{\partial Z} = \mathrm{i}|U|^2 U - \left(\frac{\mathrm{i}}{\pi}\frac{\partial}{\partial T}\right)^n U - \mathrm{i}\Delta U - U + \left[\mathrm{i}C^* U^2 + \mathrm{i}C^2 U^* + \mathrm{i}2C|U|^2\right] + \mathrm{i}2|C|^2 U. \tag{S6}$$

For bright solitons, $C$ is the solution on the lower branch of the bi-stability curve [S4]-[S6]. When searching for stationary solitons, solving Eqs. (S5) and (S6) in sequence is intrinsically equivalent to solving Eq. (S2).

## 1.2 Eigen functions of combined spectral filtering and self-phase modulation

### 1.2.1 Evolution with filter order

The eigenfunctions of combined spectral filtering and self-phase modulation (SPM) obey

$$\mathrm{i}|U_\mathrm{e}|^2 U_\mathrm{e} - \left(\frac{\mathrm{i}}{\pi}\frac{\partial}{\partial T}\right)^n U_\mathrm{e} = (\mathrm{i}\xi + \lambda)U_\mathrm{e}, \tag{S7}$$

where $\lambda$ and $\xi$ represent the real and imaginary parts of the eigenvalue respectively. It is noted that $\lambda$ and $\xi$ are correlated. For a given $\xi$, the eigenfunction $U_\mathrm{e}$ and the real eigenvalue $\lambda$ can be obtained by numerically solving Eq. (S7) with the Newton-Rapson method. Figure S1 shows the evolution of $U_\mathrm{e}$ and $\lambda$ with the filter order when $\xi = 1$. It is found that $\lambda$ is always negative, representing an overall amplitude loss induced by SPM in combination with spectral filtering. Remarkably, $|\lambda|$ decreases with the increase of filter order $n$. When the filter order $n \to \infty$, we have $|\lambda| \to 0$.

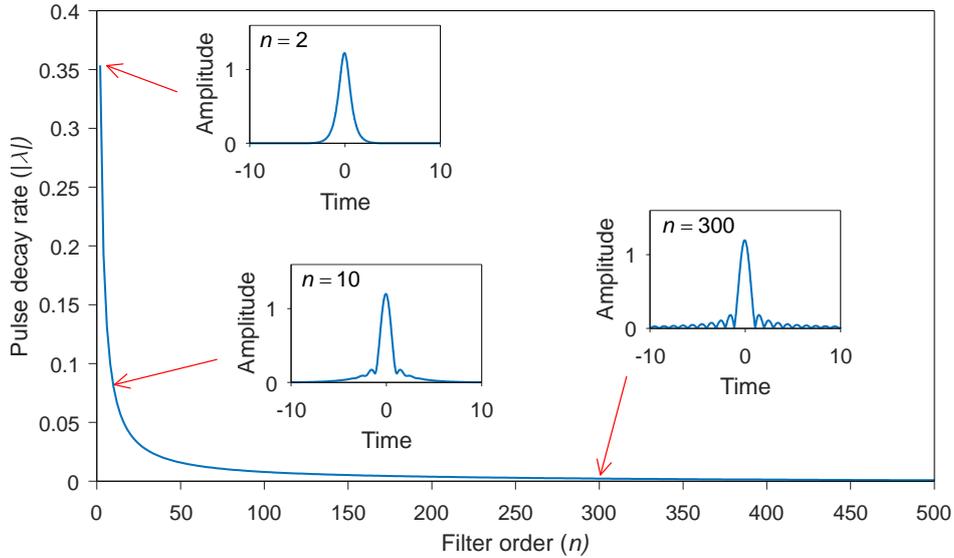

**Fig. S1 | Evolution of pulse shape and decay rate with the filter order.** The imaginary eigenvalue $\xi = 1$.

Although the eigenfunctions do not necessarily constitute stable solitons, they provide important insights for the dispersion-less dissipative solitons (as is shown by the comparison in Fig. 1c of the main paper) and may be regarded as a kind of soliton "kernel". One particularly interesting case is when $|\lambda| \to 0$ with $n \to \infty$. Note that the spectral loss can be expressed in the frequency domain as

$$H(\nu) = (2\nu)^n. \tag{S8}$$

Therefore, when $n \to \infty$, it turns to an ideal gate filter with a unit bandwidth

$$H(v) = \begin{cases} 0, & -1/2 \le v \le 1/2 \\ \infty, & \text{others} \end{cases}. \tag{S9}$$

The vanishment of $|\lambda|$ implies that in this limiting case, an energy-conserved balance can be achieved between SPM and the gate bandpass filtering. The resulting waveform is close to a Nyquist pulse with a fully confined spectrum.

### 1.2.2 Scaling law

The scaling law for the eigenfunctions can be easily checked as

$$\xi \to \eta\xi,\ \lambda \to \eta\lambda,\ U_e \to \eta^{1/2}U_e,\ T \to T\eta^{1/n}, \tag{S10}$$

where $\eta$ is a positive number. The pulse energy is then

$$E = \int \eta \left| U_e\left(\eta^{1/n}T\right) \right|^2 dT = \eta^{1-1/n} E_0, \tag{S11}$$

where $E_0$ is the energy of $U_e(T)$, given by

$$E_0 = \int \left| U_e(T) \right|^2 dT. \tag{S12}$$

And the pulse width is

$$W = \eta^{-1/n} W_0, \tag{S13}$$

where $W_0$ is the pulse width of $U_e(T)$. The relation between $E$ and $W$ is thus given by

$$E = E_0 W_0^{n-1} \left(\frac{1}{W}\right)^{n-1} = C_n \left(\frac{1}{W}\right)^{n-1}, \tag{S14}$$

where $C_n = E_0 W_0^{n-1}$ is a constant depending on the filter order. The pulse energy scales with the $(n-1)$-th power of the inverse pulse duration. Again, the interesting condition is when $n \to \infty$. The pulse width will be nearly unchanged when the pulse energy increases (also see Eq. (S10)), as a result of a spectrum that is fully confined by an ideal gate filter.

### 1.2.3 Nyquist-pulse-like solution

Equation (S7) is not integrable, thus no precise analytic solution can be obtained for $U_e$. In the next, an approximate solution is constructed for the Nyquist-pulse-like eigenfunction with $n \to \infty$. Intuitively, we assume the spectral envelop when $\xi = 1$ is given by

$$\mathbb{U}_e = \mathcal{F}\{U_e\} = a\left[\frac{1+\cos(bv)}{2}\right], \tag{S15}$$

where $\mathcal{F}$ represents Fourier transform, $a$ and $b$ are constants. Transforming Eq. (S7) with $\xi = 1$

and $n \to \infty$ to the frequency domain yields

$$i(\mathbb{U}_e \otimes \mathbb{U}_e^*) \otimes \mathbb{U}_e - H\mathbb{U}_e = i\mathbb{U}_e, \tag{S16}$$

where $H$ is the ideal gate filter (i.e., Eq. (S9)). The values of $a$ and $b$ may be found analytically by using the variational method. A much simpler approach is just fitting the numerical results with Eq. (S15). The retrieved parameters are $a = 1.382$ and $b = 2.669$. The time-domain pulse shape is then calculated by inverse Fourier transform as follows

$$U_e = \mathcal{F}^{-1}\{\mathbb{U}_e\} = \frac{a\sin(\pi T)}{\pi T} + \frac{2ab\sin\left(\frac{b}{2}\right)\cos(\pi T) - 4\pi a\cos\left(\frac{b}{2}\right)T\sin(\pi T)}{b^2 - 4\pi^2 T^2}. \tag{S17}$$

Figure S2 compares the numerical results with the calculated results based on Eqs. (S15) and (S17), showing very good agreement. According to the scaling law given by Eq. (S10), the eigenfunction spectrum for arbitrary $\xi$ can be written as

$$\mathbb{U}_e = \sqrt{\xi}a\left[\frac{1+\cos(bv)}{2}\right]. \tag{S18}$$

The results above are obtained with the normalized equation. For any practical experimental configuration, the soliton parameters can be easily obtained by performing denormalization.

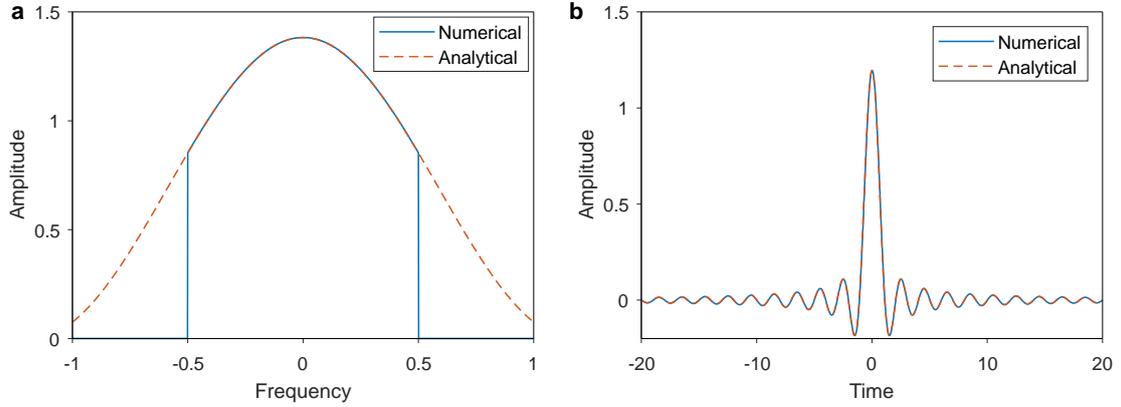

**Fig. S2 | Comparison between numerical and approximate analytical results for the Nyquist-pulse-like eigenfunction. a,** Spectrum. **b,** Pulse shape.

### 1.3 Nyquist cavity solitons

In the next, based on Eq. (S15), we derive an approximate solution for the Nyquist-pulse-like solitons in coherently driven cavities. The procedure is similar to perturbation analysis for conventional dispersion-driven solitons [S4]-[S6], but performed in the frequency domain because the time-domain pulse shape of Eq. (S17) is complicated. We first transform Eq. (S6) to the frequency domain as follows

$$\frac{\partial \mathbb{U}}{\partial Z} = i(\mathbb{U} \otimes \mathbb{U}^*) \otimes \mathbb{U} - H\mathbb{U} - i\Delta\mathbb{U} - \mathbb{U} + \left[iC^*\mathbb{U} \otimes \mathbb{U} + iC^2\mathbb{U}^* + i2C\mathbb{U} \otimes \mathbb{U}^*\right] + i2|C|^2\mathbb{U}. \tag{S19}$$

Suppose the soliton spectrum can be written as

$$\mathbb{U} = \eta \left[ \frac{1+\cos(bv)}{2} \right] e^{i\phi} = \eta Q e^{i\phi} \quad \text{with } -1/2 < v < 1/2, \tag{S20}$$

where $\eta$ and $\phi$ are spectral amplitude and phase respectively. For convenience, we use the simplified denotation $Q = [1+\cos(bv)]/2$. Substituting Eq. (S20) into Eq. (S19) and noticing that

$$i(\mathbb{U} \otimes \mathbb{U}^*) \otimes \mathbb{U} - H\mathbb{U} = i\frac{\eta^2}{a^2}\mathbb{U}, \tag{S21}$$

we will have

$$\frac{\partial \eta}{\partial Z}Qe^{i\phi} + i\frac{\partial \phi}{\partial Z}\eta Qe^{i\phi} = i\frac{\eta^2}{a^2}\eta Qe^{i\phi} - i\Delta\eta Qe^{i\phi} - \eta Qe^{i\phi} \\ + \left[ iC^*\eta^2 e^{i2\phi}Q \otimes Q + iC^2\eta Qe^{-i\phi} + i2C\eta^2 Q \otimes Q \right] + i2|C|^2 \eta Qe^{i\phi} \tag{S22}$$

The terms enclosed in square brackets represent parametric mixing between the soliton and the homogenous background. Under the small-perturbation assumption, one balance is achieved between SPM and gate spectral filtering, while the other is achieved between parametric gain and uniform loss. The frequency-dependence of parametric gain can then be neglected in approximate analysis. Thus, we assume $Q \otimes Q \approx kQ$ where the constant $k = 0.687$ can be retrieved with numerical curve fitting. By separating the real and imagery terms of Eq. (S22), we obtain the following coupled equations

$$\frac{\partial \eta}{\partial Z} = -\eta + k|C|\eta^2 \sin(\phi - \phi_C) + |C|^2 \eta \sin[2(\phi - \phi_C)], \tag{S23a}$$

$$\frac{\partial \phi}{\partial Z} = -\Delta + \frac{\eta^2}{a^2} + 3k|C|\eta\cos(\phi - \phi_C) + |C|^2 \cos[2(\phi - \phi_C)] + 2|C|^2, \tag{S23b}$$

where $\phi_C = \arg(C)$.

When the phase detuning $\Delta \gg 1$, the lower-branch homogenous solution $C \approx -iS/\Delta$. Further neglecting the higher-order terms with $|C|^2$, we get

$$\frac{\partial \eta}{\partial \xi} \approx -\eta + \frac{kS}{\Delta}\eta^2 \cos\phi, \tag{S24a}$$

$$\frac{\partial \phi}{\partial \xi} \approx -\Delta + \frac{1}{a^2}\eta^2 - \frac{3kS}{\Delta}\eta\sin\phi. \tag{S24b}$$

A fix-point solution of Eq. (S24) reads

$$\eta \approx a\sqrt{\Delta}, \tag{S25a}$$

$$\phi \approx \arccos\left(\frac{\sqrt{\Delta}}{akS}\right). \tag{S25b}$$

A comparison between the numerical results and the approximate analytical results is shown in Fig. S3, showing relatively good agreement. The parameters are $\Delta = 30$ and $S^2 = 60$.

From Eq. (S25a), we find that the soliton phase shifting rate (the imaginary eigenvalue $\xi = \eta^2/a^2$) is clamped by the pump-cavity phase detuning ($\Delta$), i.e., $\xi = \Delta$. This conclusion is identical to that for the conventional cavity solitons [S6], and can be easily understood by noticing the fact that the soliton phase-synchronized with the external pump will get the maximum gain when it is coherently combined with the pump field. It is noted that the analytical approximation by Eq. (S25) is more accurate with lower pump intensity under the premise of soliton sustainment. With the increase of pump power, the cavity soliton will show slightly increasing distortions in comparison to the eigenfunction, as is indicated by Fig. 5 of the main paper.

The soliton stability can be investigated according to Eq. (24). Figure S4 shows one simulated example in which the initial soliton parameters deviate from the stationary solution. With the increase of propagation distance, the soliton finally converges to the fixed point.

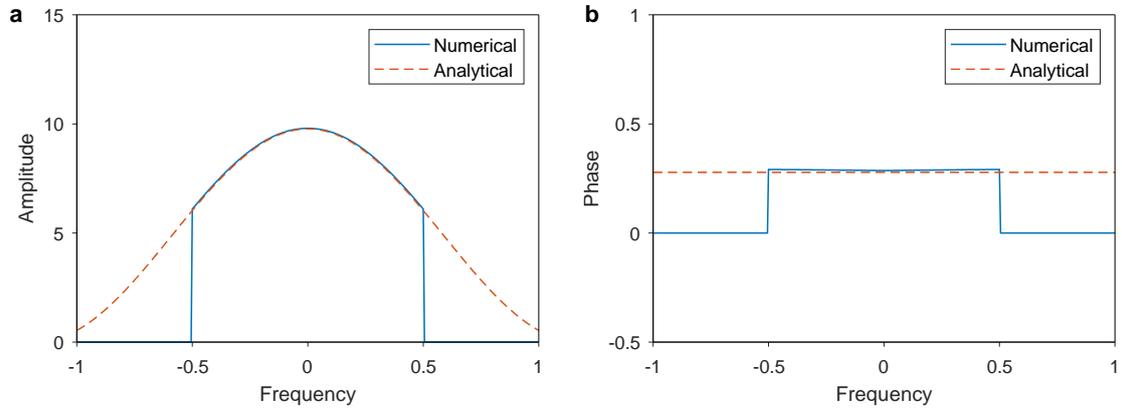

**Fig. S3 | Comparison between numerical and approximate analytical results for the Nyquist cavity soliton. a,** Spectral amplitude. **b,** Spectral phase. The simulation parameters are $\Delta = 30$ and $S^2 = 60$.

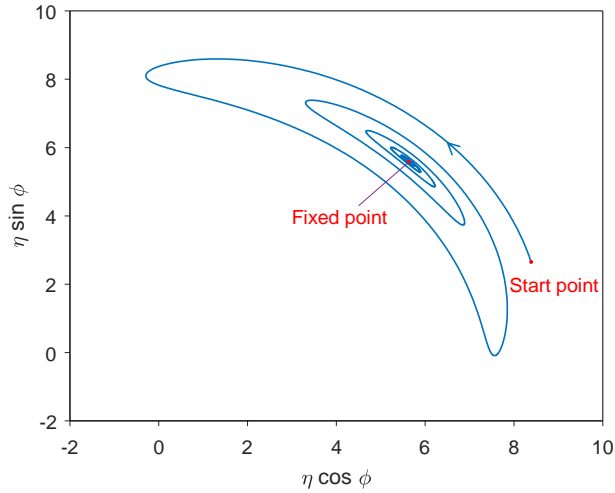

**Fig. S4 | Dynamic evolution of Nyquist cavity soliton.** The simulation parameters are $\Delta = 30$ and $S^2 = 60$.

## 1.4 Nyquist soliton molecules

Depending the initial field (e.g., when the pulse duration is much wider than the transform-limited value), stable soliton molecules composed of multiple closely bound pulses can also be observed in simulations. These Nyquist soliton molecules are also related to the eigenfunctions of combined SPM and gate spectral filtering. Figure S5 shows the numerical results simulated based on Eqs. (S5)-(S7). The parameters are $\xi = \Delta = 50$ and $S^2 = 60$. Figure 4 of the main paper shows experimental evidences of Nyquist soliton molecules and their transitioning in a pulse pumped fiber ring cavity.

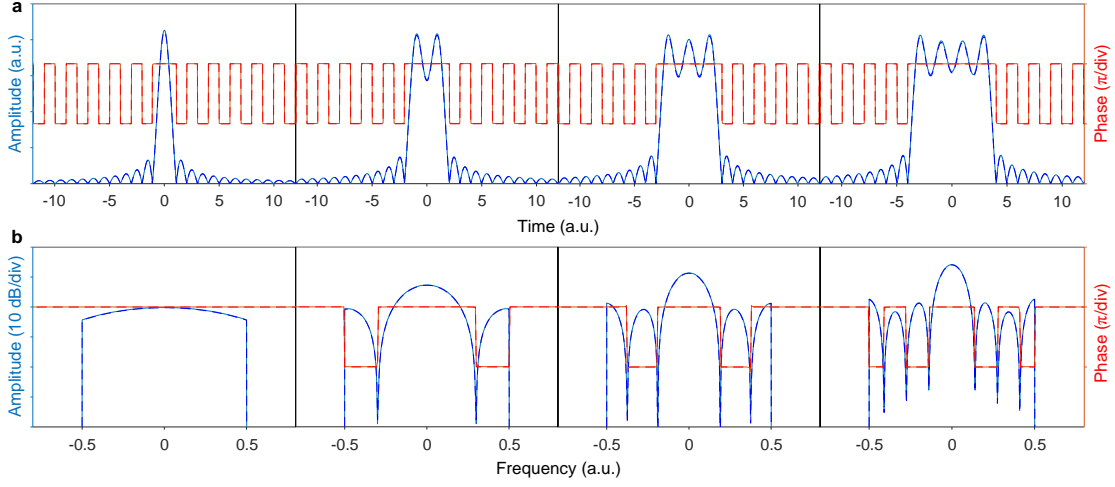

**Fig. S5 | Compact Nyquist soliton molecules. a,** Pulse shape. **b,** Spectrum. From left to right, the molecules contain 1, 2, 3 and 4 solitons, respectively. Solid: cavity soliton; dash: Kerr-and-filter eigen function. The solid and dash lines are overlaid and visually indistinguishable. The simulation parameters are $\xi = \Delta = 50$ and $S^2 = 60$.

## 1.5 Pulse pumped equation for the fiber cavity

When the Kerr cavity is pumped by optical pulses, a new term should be introduced to account for the desynchronization between the cavity soliton and the external pump pulse. The mean-field equation including the second-order dispersion reads

$$\frac{\partial A}{\partial z} = -(\alpha + i\delta)A + i\gamma |A|^2 A - \rho\left(i\frac{\partial}{\partial t}\right)^n A - i\frac{\beta_2}{2}\frac{\partial^2}{\partial t^2} A - \zeta\frac{\partial}{\partial t} + \kappa A_p \quad (S26)$$

where $\beta_2$ is the second-order dispersion; $\zeta = d/[v_g(1+d)]$ is the desynchronization induced temporal drifting; and $d = (R_p - R_c)/R_c$ represents the relative difference between the pump pulse repetition rate $R_p$ and the cavity free spectral range $R_c$; $v_g$ is the soliton group velocity.

For the simulation results shown in Figs. (3), (4) and (6) of the main paper, a simplified model is employed in which each effect is considered separately by a lumped term in one roundtrip. The field at the end of the (m+1)-th round is related to that after the m-th round by

$$A_{m+1} = \sqrt{(1-\theta)(1-\alpha_L)}e^{-i(\delta_L + \gamma P_m L)} F A_m + \sqrt{\theta} A_p \quad (S27)$$

where $A$ is the field amplitude normalized such that $P = |A|^2$ represents the field power; $\theta$ is the pump-cavity power coupling ratio; $\alpha_L$ is the uniform roundtrip power loss; $\delta_L$ is the pump-cavity phase detuning; $\gamma$ is the average Kerr coefficient; $L$ is the roundtrip length; and $A_p$ is the pump field. The operator $F$ accounts for the effects that can be easily applied in the frequency domain, including the frequency-dependent filtering loss, the group velocity dispersion, and the pump-cavity desynchronization; i.e.

$$\mathcal{F}\{FA_m\} = H_T \mathbb{A}_m e^{id_2\omega^2/2} e^{-i\omega\tau} \tag{S28}$$

where $\mathbb{A}_m = \mathcal{F}\{A_m\}$ is the field spectrum; $H_T$ represents the transfer function of the spectral filter; $d_2$ is the roundtrip group delay dispersion; $\tau$ is the desynchronization induced time shift between the cavity soliton and the pump pulse; and $\omega$ is the angular frequency. Note that calculating Eq. (S27) is actually equivalent to integrating Eq. (S26) with the split-step Fourier method with a step size equal to one roundtrip. The parameters are related by

$$\theta = \kappa^2 L^2, \ \alpha_L = 1 - e^{-2\alpha L}, \ \delta_L = \delta L, \ d_2 = \beta_2 L, \ \tau = \zeta L. \tag{S29}$$

## 2. Experiments

### 2.1 Fiber cavity stabilization

The detailed experimental setup is shown in Fig. S6. To maintain a stable pump-cavity frequency detuning for soliton generation, the laser frequency is first locked to the fiber cavity resonance by sending a probe light in the backward direction and using the Pound-Drever-Hall locking technique. An acousto-optic frequency shifter is then used to tune the frequency of the pump pulse in the forward direction. The polarizations of the probe and the pump are adjusted orthogonal to each other to minimize their mutual interference.

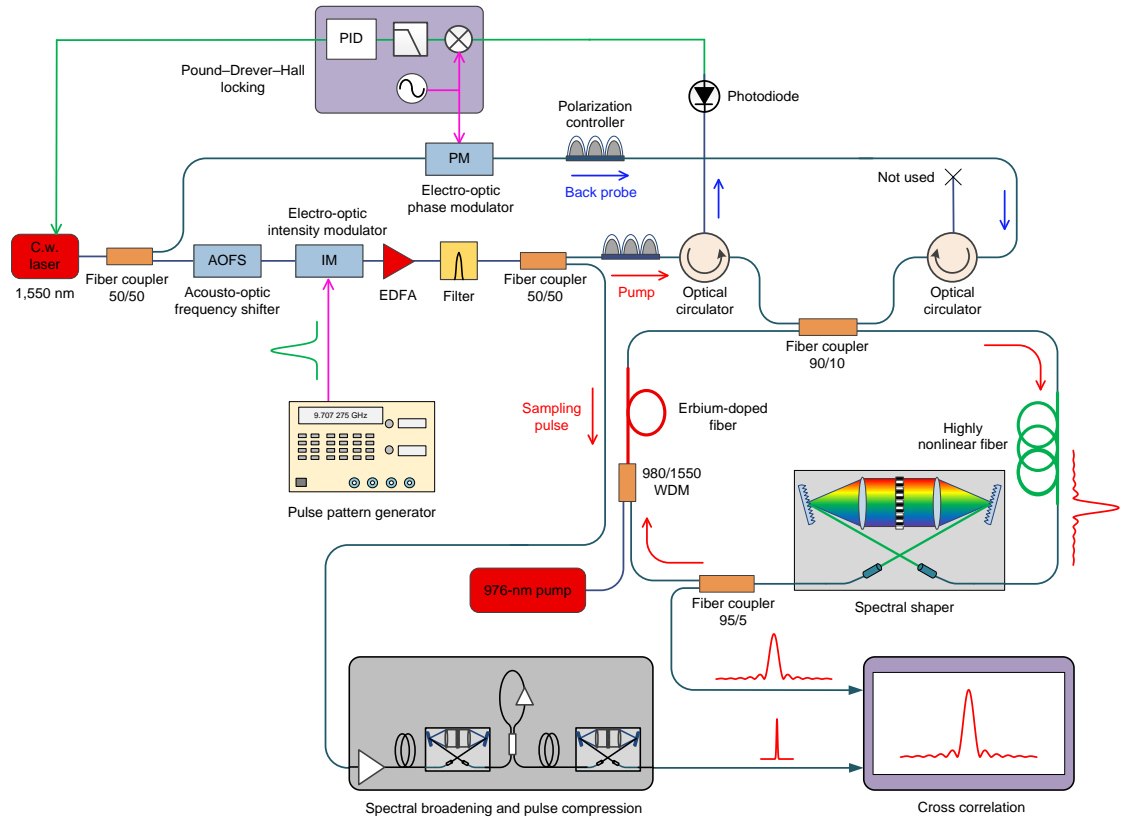

**Fig. S6 | Experimental setup for soliton generation and characterization.** PID: proportional–integral–derivative controller; EDFA: erbium-doped fiber amplifier; WDM: wavelength-division-multiplexed coupler.

## 2.2 Intensity cross correlation for soliton characterization

The intensity cross correlation setup for measuring the soliton pulse shape is shown in Fig. S7. The synchronous sampling pulse is generated through a multiple-stage spectral broadening and pulse compression procedure. A portion of the Gaussian pump pulse is first amplified to a peak power of ~200 W, and sent through a 50-m highly nonlinear fiber (HNLF). The spectrum is broadened to ~2 nm and the pulse is compressed by a spectral shaper. A nonlinear amplifying loop mirror (NALM) composed of 2-m HNLF and 0.5-m Erbium-doped fiber (LIEKKI Er-110) is used to improve the pulse quality and further broaden the spectrum to ~6 nm. After passing through another 50-m HNLF, the spectral bandwidth exceeds 100 nm. A second spectral shaper is used to select the spectrum within 1527-1567 nm and shape the pulse to a Gaussian function with a full-width-at-half-maximum of 0.2 ps. The peak power of the output pulse is ~130 W. The pulse is much narrower than the soliton to be measured, making it possible to capture the fine temporal features with a high resolution.

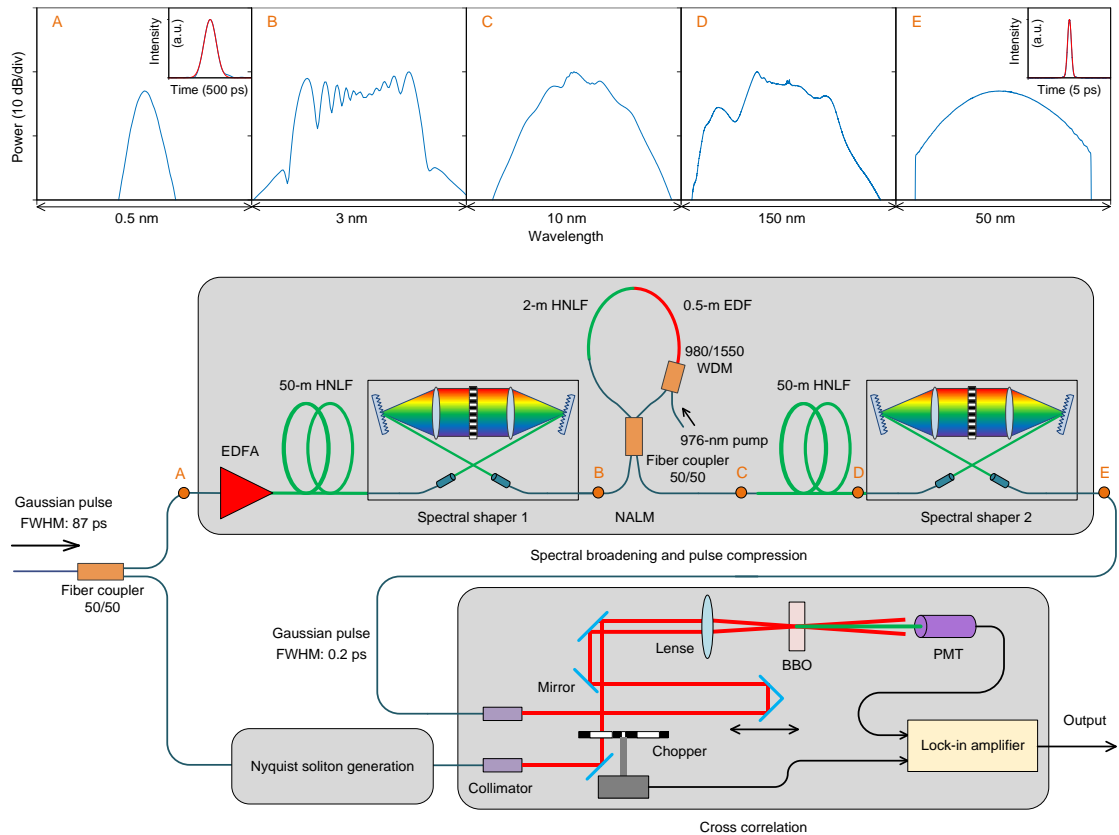

**Figure S7 | Intensity cross correlator for soliton characterization.** The insets on top show the spectra at different locations (A, B, C, D, E). The pulse waveform at point A detected by a high-speed photodiode and the autocorrelation of the compressed Gaussian pulse at point E are also shown (blue: measured; red: calculated). FWHM: full-width-at-half-maximum; EDFA: Erbium-doped fiber amplifier; HNLF: highly nonlinear fiber; EDF: Erbium-doped fiber; WDM: wavelength-division-multiplexed coupler; NALM: nonlinear amplifying loop mirror; BBO: beta barium borate crystal; PMT: photomultiplier tube.

## 2.3 Nyquist soliton transition

Figure S8 shows the full data of the soliton transition process in Fig. 4 of the main paper. With the increase of pump frequency detuning from $s_{16}$ to $s_1$, the soliton molecule pulse width decreases from ~126 ps to ~4 ps and the spectral sidelobes disappear pair by pair. A single Nyquist soliton with a smooth spectrum is finally formed.

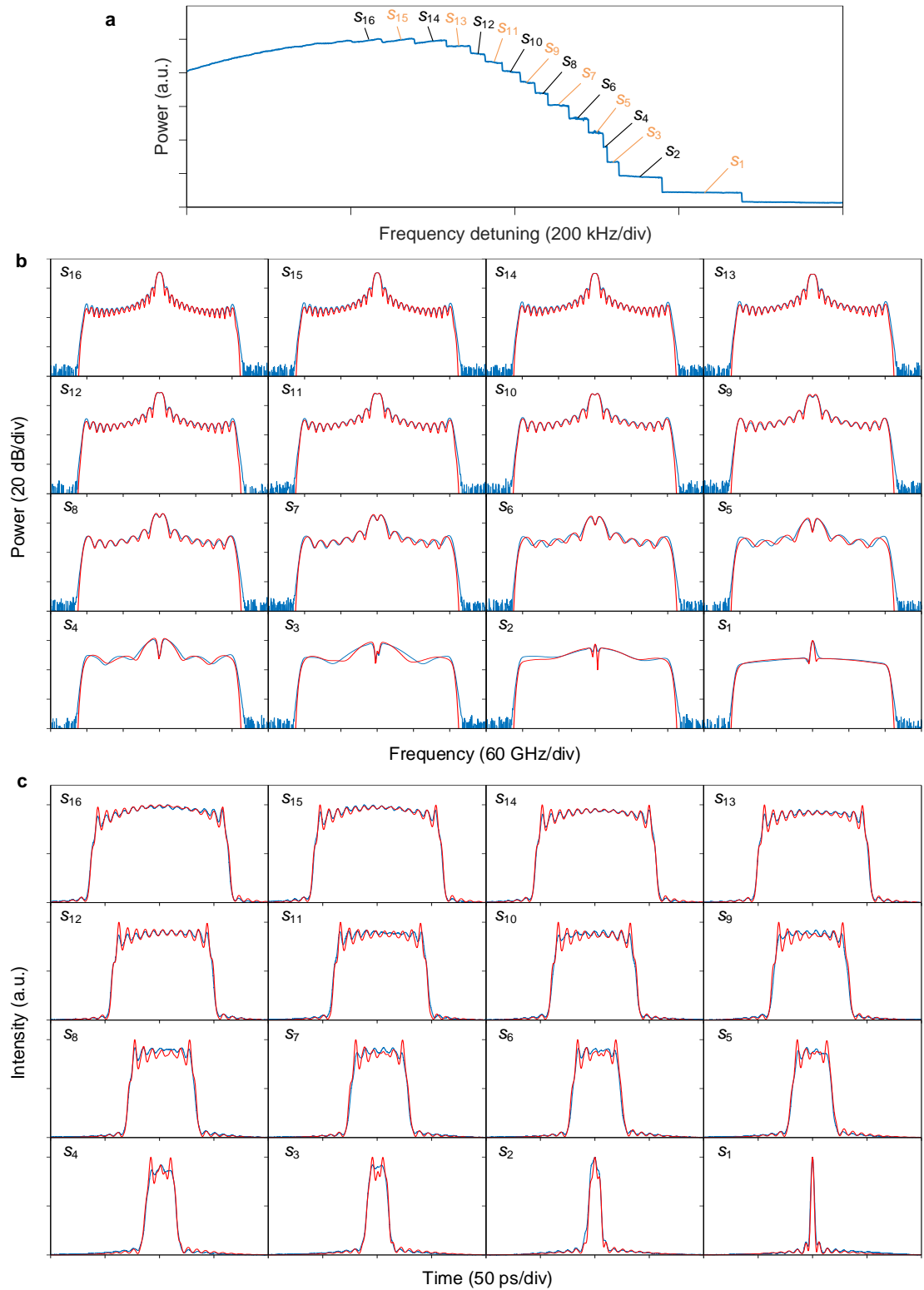

**Figure S8 | Nyquist soliton transition under pulsed pumping. a,** Intracavity power versus pump frequency detuning. **b,c,** Spectrum and pulse shape at the different steps marked in **a** ($s_{16}$ to $s_1$). Blue: measured; red: simulated.

## 2.4 Effect of pump-cavity desynchronization

It is found that the soliton transition process is affected by the desynchronization between the pump pulse and the cavity. Figure S9a shows multiple intracavity power traces measured for slightly different pump pulse repetition rates. The simulation results are shown in Fig. S9b. The measured filter transfer function shown in Fig. 2c of the main paper is employed in the simulation. The desynchronization parameter (i.e., relative difference between the pump pulse repetition rate and the cavity FSR) varies between $\pm 3\times 10^{-4}$. The widest single-soliton step is achieved when the desynchronization is slightly negative (of order of $10^{-5}$). When the desynchronization magnitude gets larger, the single-soliton region becomes narrower and may even disappear. For large desynchronization, the intracavity field drops to the lower-branch homogenous state before reaching the single-soliton state. Simulations show that the soliton transition behavior may also be affected by the pump pulse shape as well as the asymmetry of the filter function.

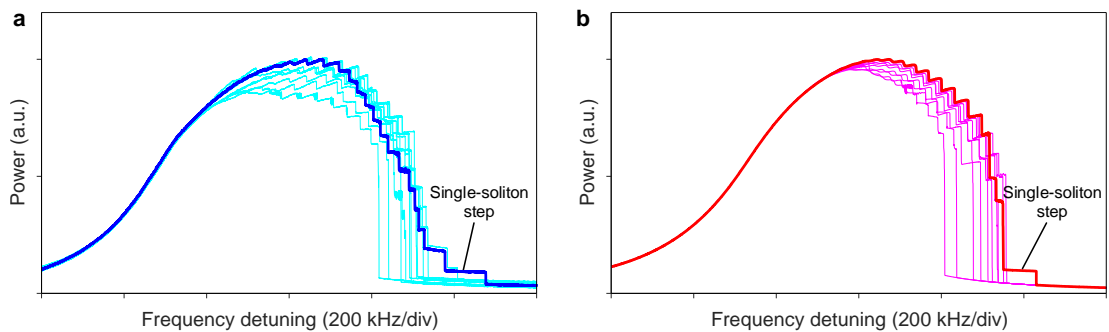

**Figure S9 | Multiple intracavity power traces with slightly different pump repetition rates. a,** Measured. **b,** Simulated. One trace showing single soliton generation is highlighted.

## 2.5 Spectral evolution in one roundtrip

Figure S10 shows the soliton spectra measured before and after the spectral shaper. The uniform insertion loss of the spectral shaper has been subtracted. No significant change can be observed. The flat pedestal in the spectrum afore spectral shaper is mainly attributed to the amplifier spontaneous emission noise.

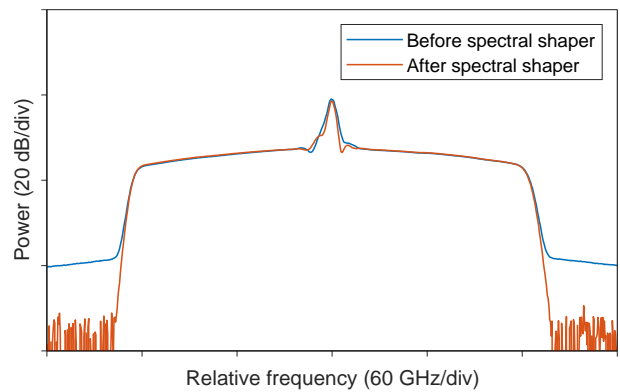

**Figure S10 | Soliton spectra before and after the spectral shaper.**

**References for supplementary information**